\documentclass[a4paper,fleqn,usenatbib]{mnras}

\usepackage{graphicx}
\usepackage[percent]{overpic}
\usepackage{lipsum}
\usepackage{multirow}
\usepackage{color}
\usepackage{rotating}
\usepackage{mwe,tikz}
\usepackage[percent]{overpic}
\usepackage{mathtools}
\usepackage{physics}
\usepackage{amsmath}
\usepackage[utf8x]{inputenc}
\usepackage{hyperref}
\usepackage{longtable}
\usepackage{wasysym}
\hypersetup{
    colorlinks=true,
    linkcolor=blue,
    filecolor=magenta,      
    urlcolor=blue,
}
\usepackage{soul}
\usepackage{pdflscape}
\usepackage{longtable}

\newcommand{\msun}{${\rm M_{\sun}}$}

\def\ltsima{$\; \buildrel < \over \sim \;$}
\def\simlt{\lower.5ex\hbox{\ltsima}}
\def\gtsima{$\; \buildrel > \over \sim \;$}
\def\simgt{\lower.5ex\hbox{\gtsima}}
\def\kms{{\rm\,km\,s^{-1}}}
\def\pc{{\rm\,pc}}
\def\kpc{{\rm\,kpc}}

\def\msun{{\rm\,M_\odot}}


\def\deg{^\circ}

\def\Gyr{{\rm\,Gyr}}

\def\masyr{{\rm\,mas \, yr^{-1}}}

\def\ltsima{$\; \buildrel < \over \sim \;$}
\def\gtsima{$\; \buildrel > \over \sim \;$}

\title[Constraining the MW halo potential]{Constraining the Milky Way Halo Potential\\
with the GD-1 stellar stream}

\author[Malhan \& Ibata]{
Khyati Malhan,$^{1,2}$\thanks{E-mail: khyati.malhan@fysik.su.se}
Rodrigo A. Ibata$^{2}$\thanks{E-mail: rodrigo.ibata@astro.unistra.fr}
\\
$^{1}$The  Oskar  Klein  Centre  for  Cosmoparticle  Physics,  Department  of Physics,  Stockholm  University,  AlbaNova,  10691  Stockholm,  Sweden\\
$^{2}$Universit\'e de Strasbourg, CNRS, Observatoire Astronomique de Strasbourg, UMR 7550, F-67000 Strasbourg, France\\
}

\date{Accepted 2019 April 9. Received 2019 April 4; in original form 2018 July 13}

\pubyear{2019}

\begin{document}
\label{firstpage}
\pagerange{\pageref{firstpage}--\pageref{lastpage}}
\maketitle

\def\MWmass{2.5\pm0.2 \times10^{11} \msun}
\def\Vcirccalc{244\pm4\kms}
\def\Vcirccalcwunc{244\kms}
\def\Vcirccross{245\kms}
\def\qrhocalc{0.82^{+0.25}_{-0.13}}
\def\qrhocalcwunc{0.82}
\def\qrhocross{0.86}
\def\veldisp{2.30\kms}
\def\veldispunitless{2.30}
\def\FeHcalc{-2.24 \pm 0.21}
\def\FeHnounc{-2.24}
\def\rapo{20.8}
\def\rperi{13.9}
\def\ecc{0.20}
\def\zmax{13.0}
\def\Lzang{2954}

\begin{abstract}
We use ESA/Gaia astrometry together with SEGUE and LAMOST measurements of the GD-1 stellar stream to explore the improvement on the Galactic gravitational potential that these new data provide. Assuming a realistic universal model for the dark matter halo together with reasonable models of the baryonic components, we find that the orbital solutions for GD-1 require the circular velocity at the Solar radius to be $V_{\rm circ}(R_\odot) =\Vcirccalc$, and also that the density flattening of the dark halo is $q_{\rho}=\qrhocalc$. The corresponding Galactic mass within $20\kpc$ was estimated to be $M_{\rm MW}(<20\kpc)=\MWmass$. Moreover, Gaia's excellent proper motions also allowed us to constrain the velocity dispersion of the GD-1 stream in the direction tangential to the line of sight to be $<\veldisp$ (95\% confidence limit), confirming the extremely cold dynamical nature of this system. 
\end{abstract}

\section{Introduction}

The mass density profile and the total mass of the dark matter halo around the Milky Way galaxy are of great astrophysical and cosmological importance, but observationally these parameters have been very hard to pin down. In recent years a wide range of solutions for the spatial distribution of the dark matter halo have been found, from close to spherical \citep{Ibata2001, Kupper2015}, oblate or prolate \citep{Law_2005_Sgr, Helmi_2004_Sgr}, to triaxial \citep{LawMajewski2010}. The disparities have persisted in part due to the lack of good quality tangential velocity measurements and distance estimates of halo tracer stars used in the dynamical analyses. This situation now looks set to change thanks to the excellent ESA/Gaia data that has recently been made available to the astronomical community \citep{GaiaDR12016, GaiaDR2_2018_Brown, GaiaCollab2018HRdiagram}. 

Various methods have been proposed and employed to constrain the mass distribution of the Milky Way galaxy. These include analyses based on the rotation curve of the Galaxy \citep{Sofue2012RotationCurve}, Jeans analyses that assume dynamical equilibrium of some tracer population to constrain the gravitational force field \citep{Loebman2014_jeans,Diakogiannis2017_Jeans}, orbital analyses of the satellite galaxies \citep{Watkins2018_GCanalysis, Fritz2018_1}, and distribution function analyses \citep{Posti2018MWMass}. However, some recent studies have turned to using stellar streams as dynamical probes of the dark matter distribution \citep{Ibata2001, Helmi_2004_Sgr, Koposov2010, Kupper2015}.

\begin{figure*}
\begin{center}
\includegraphics[width=\hsize]{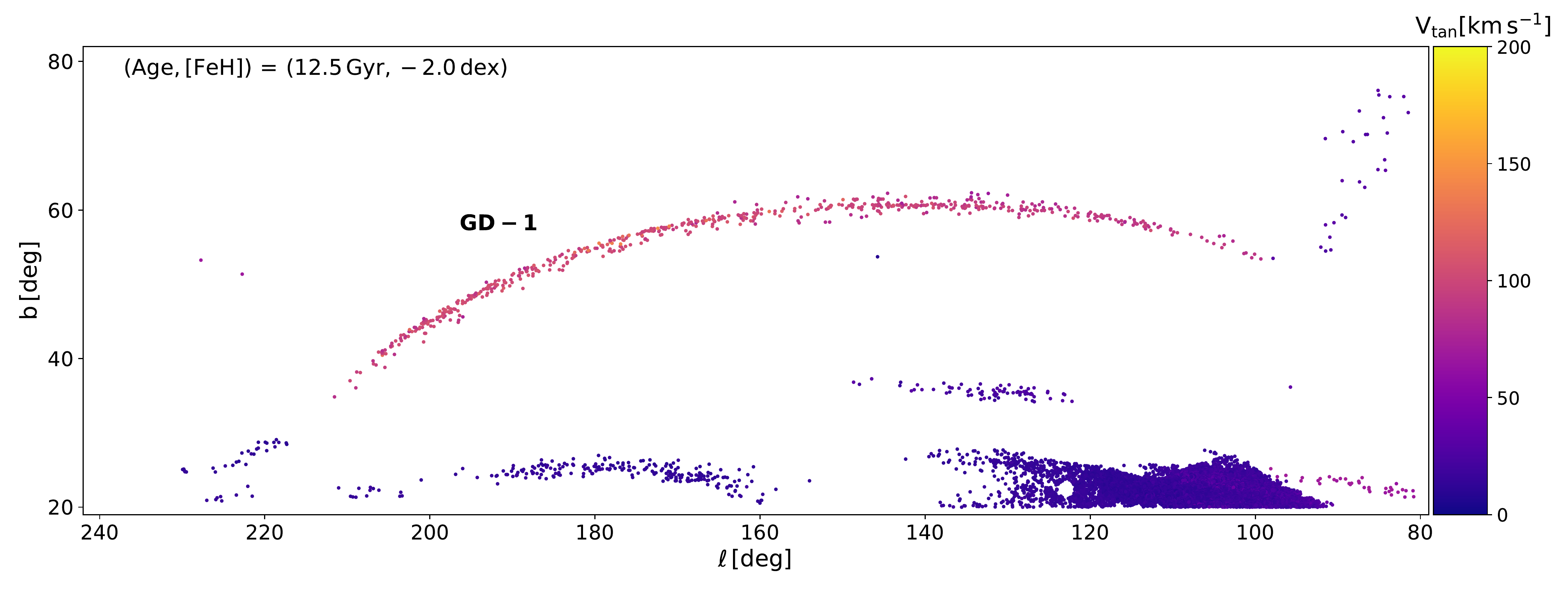}
\vspace{-0.7cm}
\end{center}
\caption{ GD-1 stream in the \texttt{STREAMFINDER} density map. The figure shows the stream detection density plot that we obtained from the \texttt{STREAMFINDER} algorithm after its application on the Gaia DR2 dataset. The map corresponds to the Stellar Population template model of (Age,[Fe/H]) = $(12.5\Gyr, -2.0)$, and shows stars with detection significance $> 10\sigma$. The points here are colored in accordance with their tangential velocities as estimated by the algorithm. The $\sim 75\deg$ long GD-1 stream stands out strikingly in this plot, which provides the basis for the sample used in our analysis.}
\label{fig:Streamfinder_density_plot}
\end{figure*}

Stellar streams are structures that are formed via the tidal disruption of globular clusters or dwarf galaxies as they orbit around their host galaxy \citep{Johnston1998, HelmiActionAgle1999}. In the low mass limit, the track traced by a stream closely delineates an orbit \citep{Dehnen2004thinorbit, EyreBinney2011}, and this orbital property can be exploited to constrain the underlying gravitational potential and the dark matter distribution \citep{Ibata2001, Koposov2010, Varghese2011, Kupper2015, Bovy2016GD1Pal5}, especially in the Milky Way where accurate measurements of the kinematics and distances of stars are available. Now with the arrival of the second data release (DR2) of the ESA/Gaia mission \citep{GaiaDR2_2018_Brown, GaiaDR2_2018_astrometry}, which has provided a huge leap in the quality of phase-space information of streams in the Milky Way, we may re-appraise the constraints on the Milky Way's potential provided by these structures.

In this contribution we make use of one of the most prominent streams in our Galaxy, the GD-1 stellar stream \citep{GrillmairGD12006}, to probe the Milky Way's gravitational potential. Situated at an intermediate Galactocentric distance of $\sim14\kpc$, this very extended pencil-line structure \citep{WhelanBonacaGD12018} is among the highest contrast streams in the Gaia dataset, making it a useful case study to assess the improvement on the Galactic gravitational potential provided by the excellent proper motions in Gaia DR2.

This paper is arranged as follows. In Section \ref{sec:Data} we discuss the data used and selections made, Section \ref{sec:Constrain_MW_potential} presents the dynamical analysis of the GD-1 stream, in Section \ref{sec:GD1_vel_Feh_Feh} we estimate some of the useful physical attributes of GD-1, and we conclude and discuss our results in Section \ref{sec:Conclusions}.

\begin{figure*}
\begin{center}
\includegraphics[width=0.9\hsize]{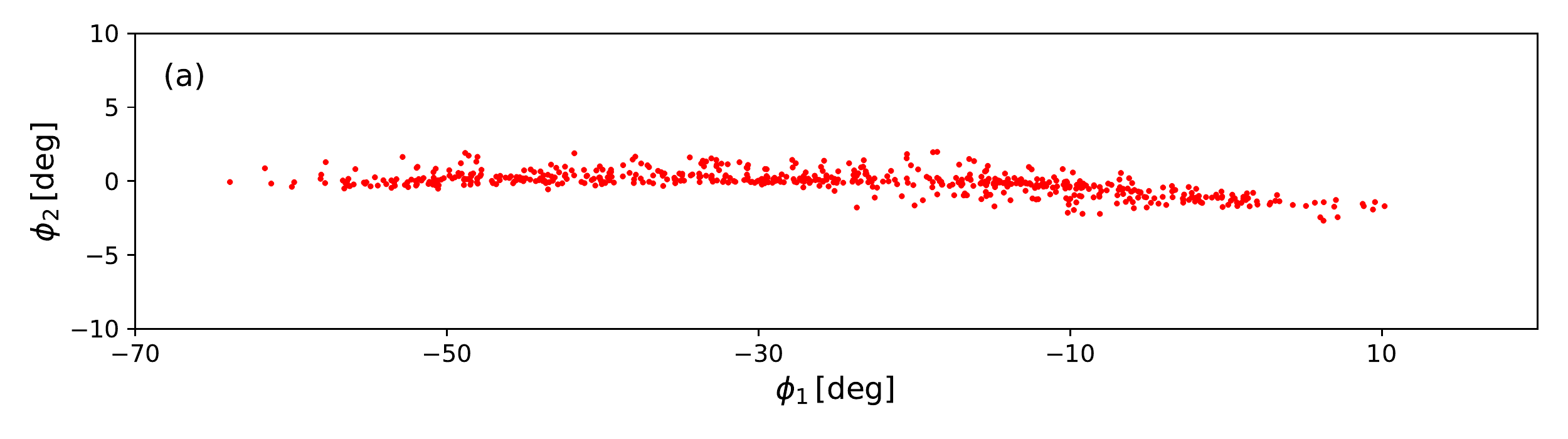}
\includegraphics[width=0.9\hsize]{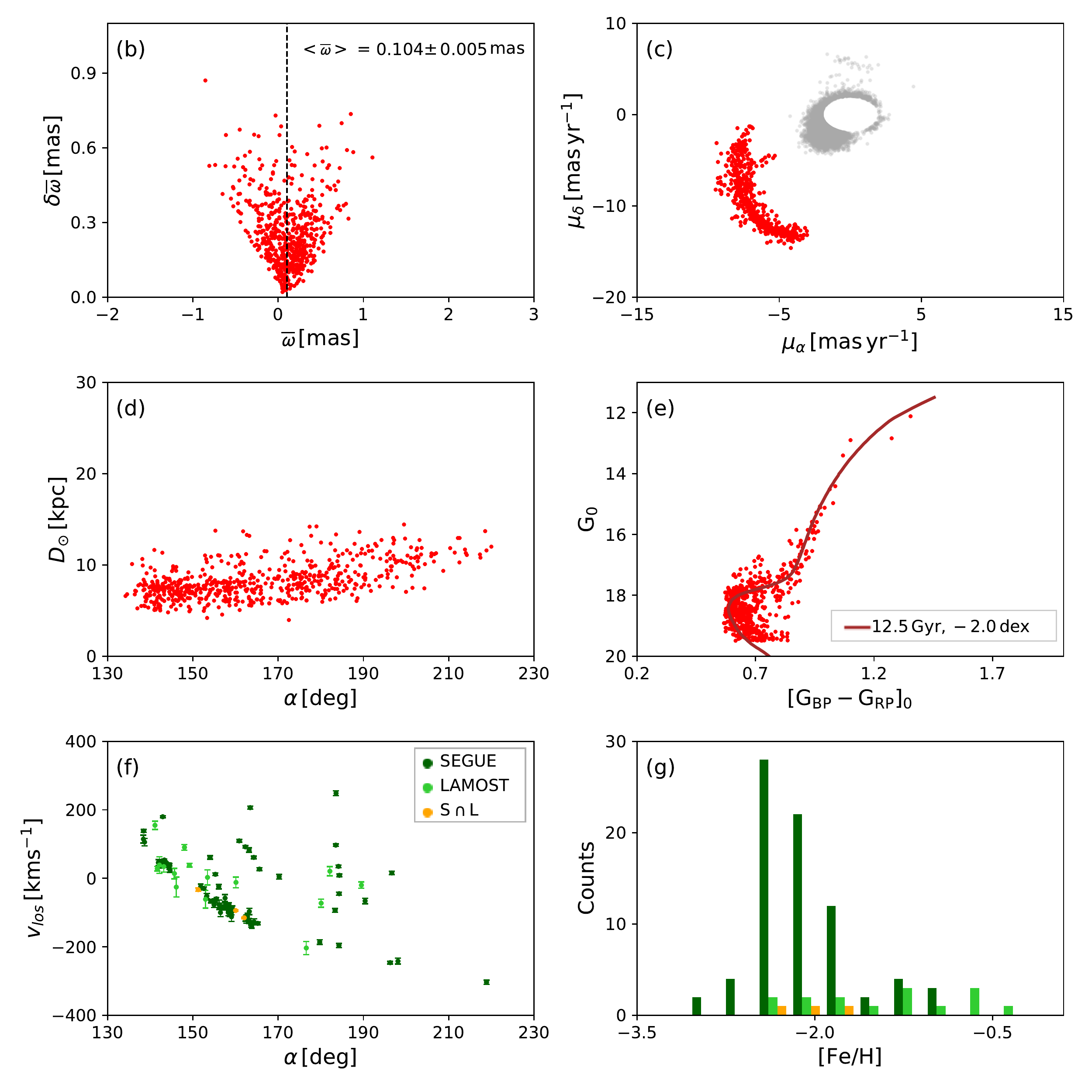}
\vspace{-0.5cm}
\end{center}
\caption{GD-1 stream phase-space map and chemistry. We extracted a $\sim10\deg$ wide region around the GD-1 stream in Figure \ref{fig:Streamfinder_density_plot} yielding $605$ stars. The parameters shown here are (a) position (in rotated frame of the GD-1 stream, following conversion from \citealt{Koposov2010}), (b) parallax, (c) proper motion, (d) \texttt{STREAMFINDER}'s heliocentric distance solutions, (e) color-magnitude, (f) $v_{\rm los}$ and (g) ${\rm [Fe/H]}$ distribution of these stars. Panels (a), (b), (c), (d) and (e) represent all the stars in the sample observed by Gaia, while (f) and (g) show the $97$ cross-matches that we found in the SEGUE and LAMOST dataset (S $\cap$ L refers to the stars that had overlapping observations between SEGUE and LAMOST). The SSP model we chose to use in the \texttt{STREAMFINDER} is shown in panel (e), shifted to account for a distance modulus of 14.6 mag (but note that there is a substantial distance gradient along the structure). The grey stars in proper motion space are the remaining stars identified as candidates by the \texttt{STREAMFINDER}.}
\label{fig:Streamfinder_GD1_stars_raw}
\end{figure*}
\begin{figure}
\begin{center}
\includegraphics[width=0.9\hsize]{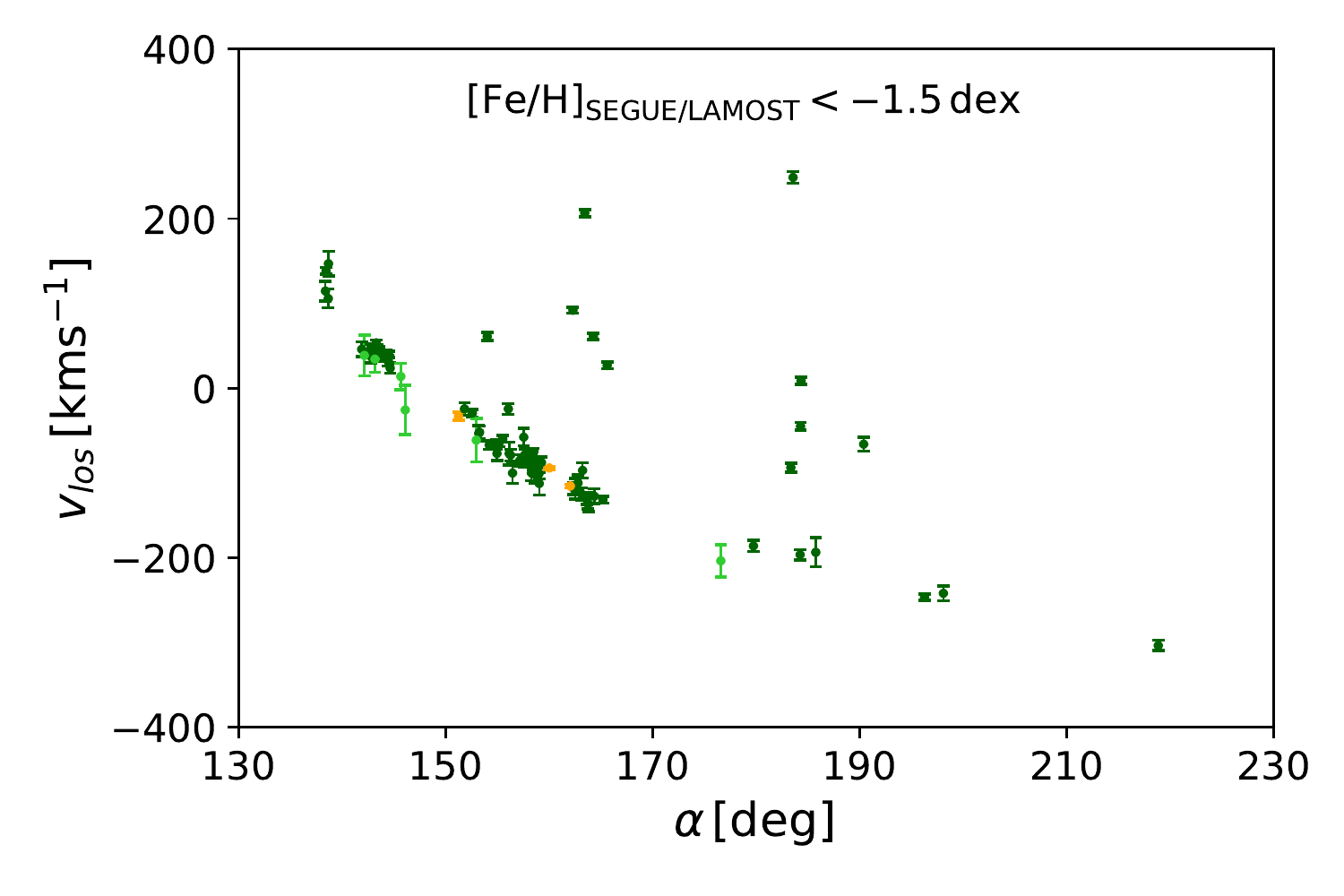}
\vspace{-0.5cm}
\end{center}
\caption{GD-1 member stars. In order to purify the GD-1 sample, presented in Figure \ref{fig:Streamfinder_GD1_stars_raw}, from contamination, we excised all the stars that possessed ${\rm [Fe/H]}>-1.5$. The reduced sample of $80$ stars is shown here. We refer to this dataset as sample-1.}
\label{fig:GD1_6D_stars_cleaned}
\end{figure}
%
\section{Data}
\label{sec:Data}

The selection of our sample of GD-1 member stars was made from the output of the \texttt{STREAMFINDER} algorithm \citep{Malhan2018_SF, Malhan_2018_PS1, Malhan_Ghostly_2018, Ibata_Norse_streams2019}, obtained from processing the Gaia DR2 dataset, after adopting a Single Stellar population (SSP) template model of ${\rm (Age,[Fe/H])} = (12.5\Gyr, -2.0)$ from the PARSEC stellar tracks library \citep{Parsec_isochrones2012}. In order to detect the GD-1 stream in particular, the algorithm was made to process only those stars in the region of sky $80\deg < \ell < 230\deg$ and $b>20\deg$, and with Heliocentric distances in the range 1 to $20\kpc$. All other algorithm parameters are identical to those described in \cite{Ibata_Norse_streams2019}. The corresponding stream map is shown in Figure \ref{fig:Streamfinder_density_plot}, where all the sources have a stream-detection significance of $>10\sigma$. This detection statistic means that at the position of each of these stars, the algorithm finds that there is a $>10\sigma$ significance for there to be a stream-like structure (with Gaussian width of $100\pc$, and $\pm 10\deg$ long) passing through the location of each of the stars, and with proper motion consistent with that of the stars. Note that this does not mean that the individual stars are stream members with $>10\sigma$ confidence.

In addition to identifying some additional stream structures, the algorithm detects the GD-1 stream that stands out strikingly from the background of contaminating stars. Although \texttt{STREAMFINDER} detects streams by looking along orbits integrated in an assumed Galactic potential model, we have shown in \cite{Malhan2018_SF} that the stream detection itself does not depend sensitively on the chosen potential model, as long as a reasonably realistic Galactic mass model is used. 

We drew a generous $\sim10\deg$ wide irregular polygon around the GD-1 structure in the map shown in Figure \ref{fig:Streamfinder_density_plot}. This selection yielded $605$ potential GD-1 stars. These stars were then cross-matched with the SEGUE  \citep{SEGUE_SDSS2009} and LAMOST \citep{LAMOST2012} datasets in order to acquire their line-of-sight velocities $(v_{\rm los})$ that are missing in Gaia DR2\footnote{Gaia DR2 measures $(v_{\rm los})$ only for the stars with $\rm{G}\simlt13$.}. A total of $97$ GD-1 candidate members yielded positive cross-matches ($82$ from SEGUE and $18$ from LAMOST) from which we obtained their $v_{\rm los}$ (we adopt the {\tt elodiervfinal} velocity measurement in SEGUE) and metallicity (${\rm [Fe/H]}$) measurements. We found that three pairs of these stars are present in both the SEGUE and LAMOST samples, and we combined the corresponding velocity and metallicity measurements (with a weighted average). Two of the $97$ cross-matched stars were found to have poorly measured ${\rm [Fe/H]}$, but we still kept them in our sample as they have well measured $v_{\rm los}$ values. This GD-1 sample is represented in Figure \ref{fig:Streamfinder_GD1_stars_raw}. We retained only these $97$ stars so as to possess a GD-1 dataset that contains complete 6D phase-space stellar information. These stars are also listed in Table \ref{tab:GD1_inventory}. 

Figure \ref{fig:Streamfinder_GD1_stars_raw} portrays GD-1 structure to be coherent in position, proper motion and colour-magnitude space. However, a glance at the $v_{\rm los}$ and the ${\rm [Fe/H]}$ distributions (panels f and g, respectively) immediately reveals the presence of contaminants in the sample, which will need to be removed or accounted for. For our kinematic analysis we impose a very conservative metallicity cut, selecting metal-poor stars with ${\rm [Fe/H]} < -1.5$ so as to retain a maximal population of GD-1 stars (we will return to the issue of measuring the metallicity of GD-1 later in Section \ref{sec:GD1_vel_Feh_Feh}). This simple metallicity cut rejects most of the outliers, as can be seen in Figure \ref{fig:GD1_6D_stars_cleaned}, resulting in a reduced sample of $80$ stars. The $v_{\rm los}$ distribution now clearly reveals a clear trend with sky position, although a few contaminants still remain. It could be tempting to simply discard the obvious outliers by hand, but we choose instead to make use of an objective algorithm which will be discussed in the next section.


%
\begin{figure*}
\begin{center}
\hbox{
\includegraphics[width=0.55\hsize]{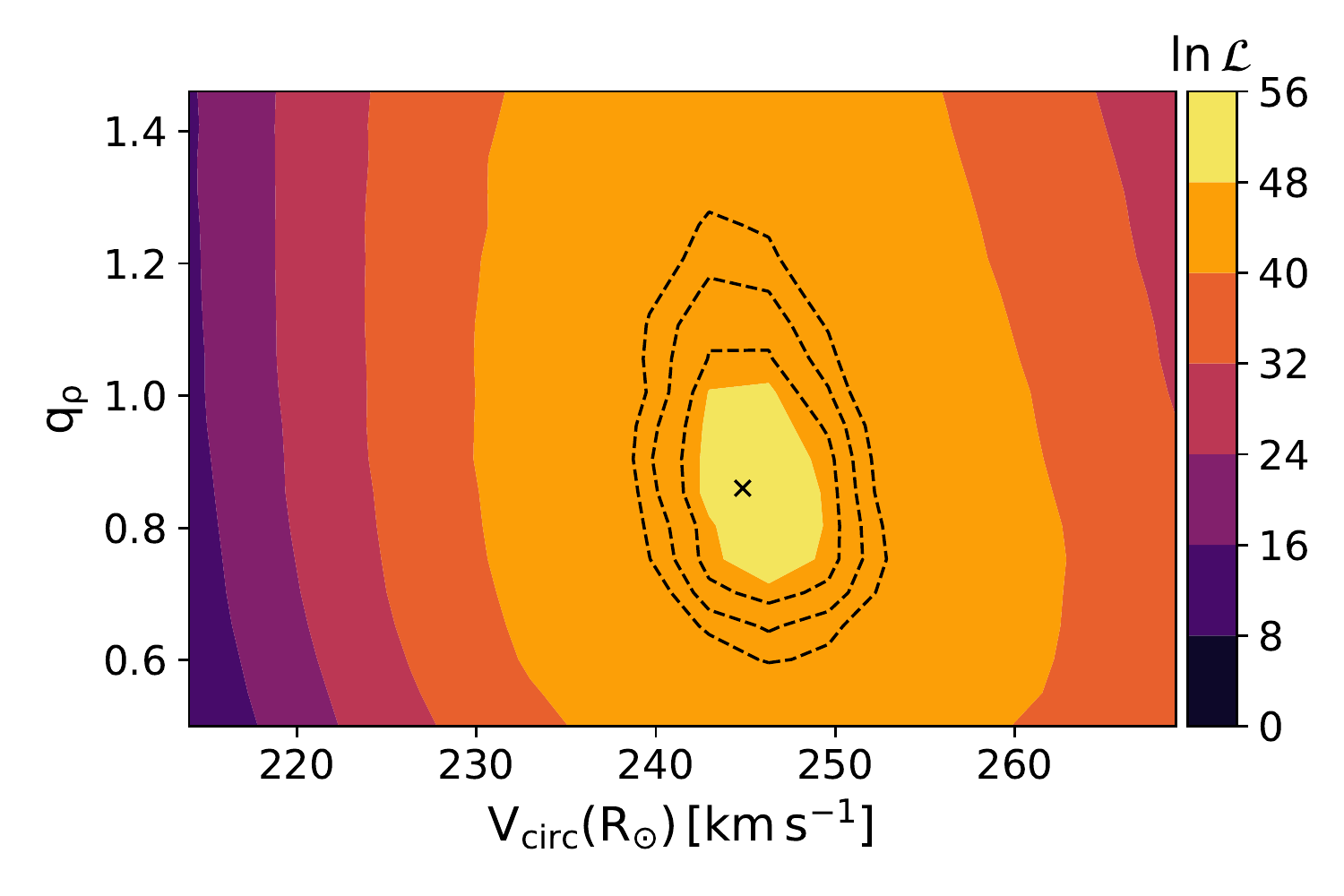}
\includegraphics[width=0.44\hsize]{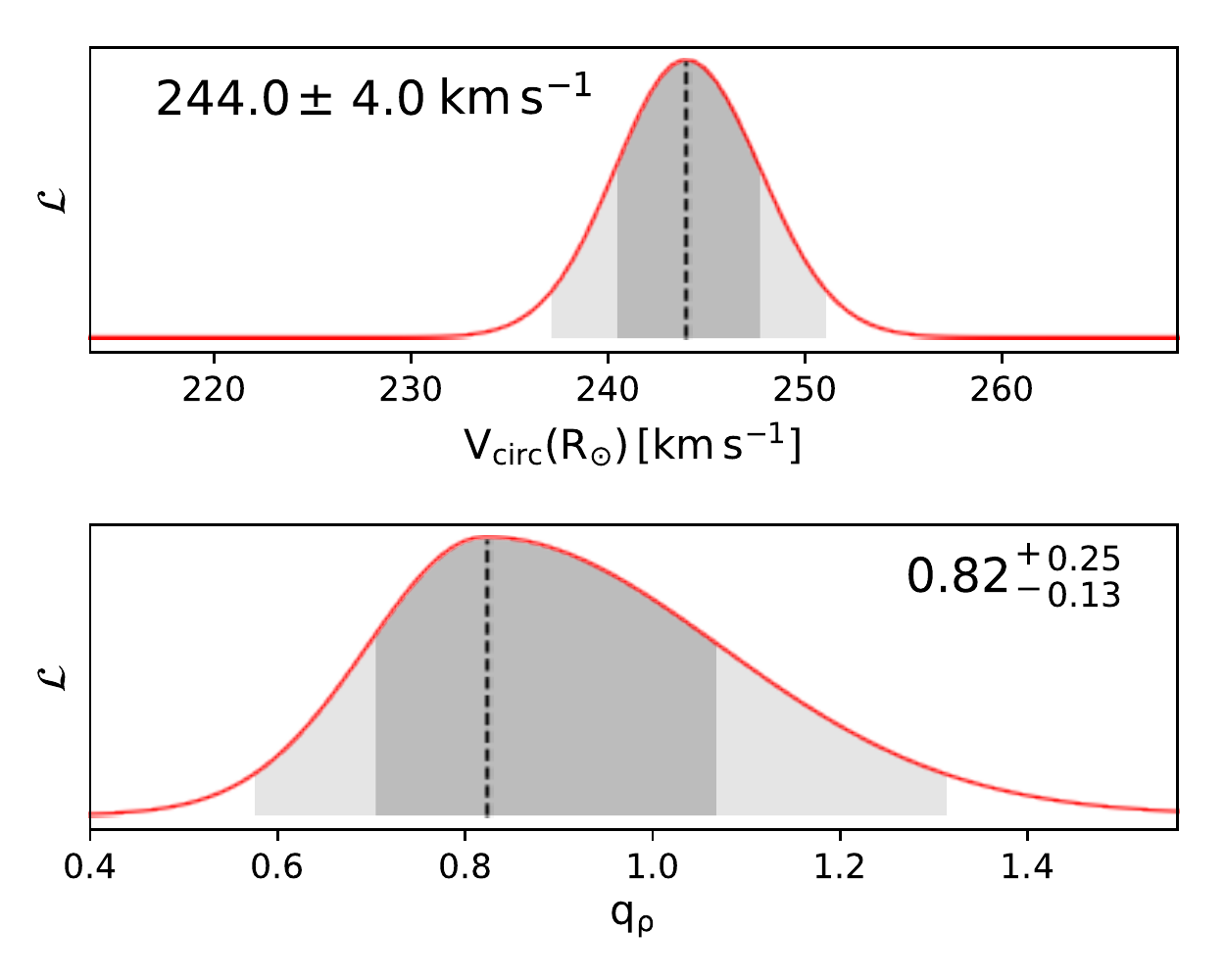}
}
\includegraphics[width=\hsize]{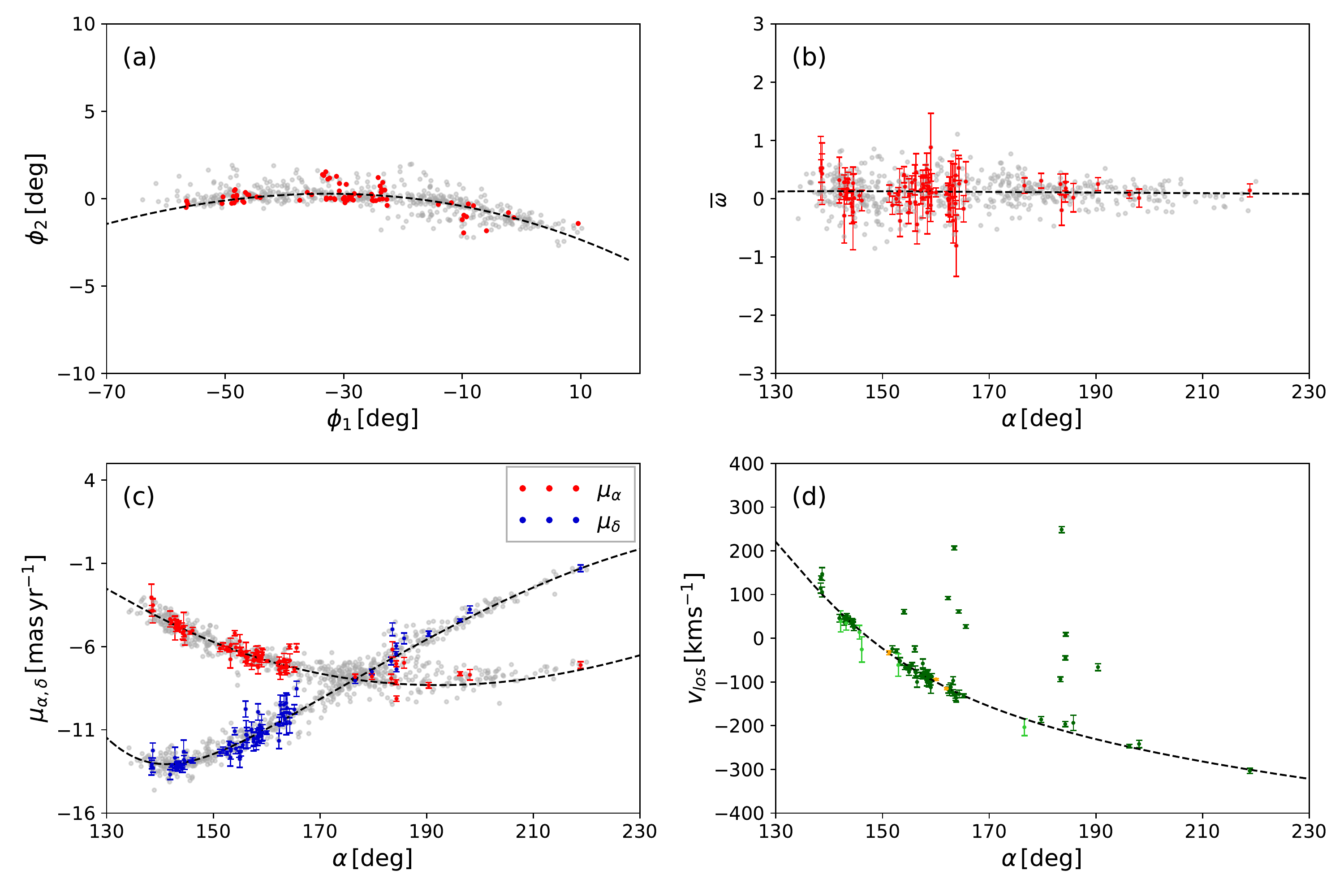}
\end{center}
\vspace{-0.5cm}
\caption{Orbit-fitting for the Milky Way halo potential. {\it Top left panel:} The contours of log-likelihood obtained from our analysis presented in Section \ref{sec:Constrain_MW_potential}. The log-likelihood surface peaks at the tuple value of $(V_{\rm circ}(R_\odot), q_{\rho})=(\Vcirccross, \qrhocross)$, corresponding to the best-fit parameters, and is marked with a cross.  The black dashed contours show the $1\sigma, 2\sigma, 3\sigma$ confidence regions. 
{\it Top right panel:} The 1D marginalised probability distributions for the parameters. This marginalisation provides the estimates $(V_{\rm circ}(R_\odot), q_{\rho})=(\Vcirccalc, \qrhocalc)$. The darker and lighter shaded areas here highlight the $1\sigma$ and $2\sigma$ confidence regions.
{\it Bottom panels:} The data-models comparison for the best-fit orbit. Sample-1 data points used for the analysis are highlighted in color, while the entire set of GD-1 stars are shown in grey. The orbit corresponding to the best fit is shown as black dashed curve.}
\label{fig:Orbit_fitting1}
\end{figure*}

\section{Constraining the Milky Way halo Potential}
\label{sec:Constrain_MW_potential}

Stellar streams of low mass progenitors closely follow orbits \citep{Dehnen2004thinorbit, EyreBinney2011} and hence their orbital properties are often exploited to constrain the underlying gravitational potential. Different methods of stream dynamical analysis have been developed, including (1) the \textit{orbit-fitting procedure} where orbits are integrated in different potential models and are then compared to stream data \citep{Koposov2010, Newberg2010OrphanFit}, (2) the \textit{N-body simulation procedure} where N-body simulation particles are compared with the data \citep{LawMajewski2010, Thomas2016}, (3) the \textit{particle-spray modelling} that models stellar tidal  streams with massless particles \citep{Varghese2011, Kupper2012Streakline}, and (4) \textit{action-angle methods} that make use of the properties of streams in action-angle space rather than in the conventional 6D phase-space \citep{EyreBinney2009, Bovy2016GD1Pal5}. 

In the cases where stellar streams vividly exhibit 2 tidal arms emerging out of the progenitor cluster at slightly different energies and angular momenta (like the Palomar 5 stream, \citealt{Rockosi2002,Ibata2016_pal5}), it is ideal to undertake a particle-spray approach or N-body simulation to allow for more realistic modelling \citep{Kupper2015, Thomas2016}. However, GD-1 is observed to be a narrow linear stream structure that lacks any obvious twin tidal arm features, and to date suggestions of the location of the progenitor's remnant are not completely convincing \citep{Boer2018, Malhan_2018_PS1, WhelanBonacaGD12018}. Therefore, given the narrow and simple structure of GD-1 (as can be seen in Figure \ref{fig:Streamfinder_GD1_stars_raw}a), we chose to model this stream with an orbit fitting procedure.

We make use of the \texttt{galpy} module \citep{BovyGalpy2015} for the purpose of setting the Galactic potential models and for the orbit integrations. We investigate a particular, but well motivated, Milky Way potential model that comprises a stellar bulge, a stellar disk and a dark-matter halo. Such a three-component parameterization is capable of reproducing the main mass components of the Milky Way. 

We model the bulge and the disk exactly as they are prescribed in \texttt{MWPotential2014} \citep{BovyGalpy2015}. The bulge is modelled as a power-law density profile (with an exponential cut-off) and is expressed as:
\begin{equation}
\rho_b (r) =\rho_{bo}\Big(\frac{r_1}{r}\Big)^{\alpha} e^{-(r/r_c)^2},
\end{equation}
with power-law exponent $\alpha=-1.8$ and cut-off radius $r_c=1.9\kpc$. The disk is represented by a Miyamoto-Nagai potential \citep{Miyamoto_disk_1975} initialized by \texttt{MiyamotoNagaiPotential} and expressed as:
\begin{equation}
\Phi_d(R, z) = - \dfrac{GM_d}{\sqrt[]{R^2+(b+\sqrt[]{z^2+c^2})^2}} \, ,
\end{equation}
setting $b$ and $c$ to the values $3.0\kpc$ and $0.28\kpc$ respectively. 

We describe the dark-matter halo by a Navarro-Frenk-White (NFW) halo model, which is well motivated by cosmological simulations \citep{NFW1997}. We use an axisymmetric NFW profile (unlike the spherical NFW profile that is used in \texttt{MWPotential2014}), given by:
\begin{equation}
\rho_h (x,y,z) = \frac{M_{vir}}{4\pi r_s^3}\frac{1}{(m/r_s)(1+m/r_s)^2},
\end{equation}
where
\begin{equation}
m = x^2 +\frac{y^2}{(b_h/a_h)^2} +\frac{z^2}{(c_h/a_h)^2} \, .
\end{equation}

The ratios between $a_h, b_h, c_h$ set the triaxiality of the dark matter halo. For the NFW halo we adopted the default values for $r_s=16.0\kpc$, following \cite{BovyGalpy2015}, and set $a_h,b_h=1$ forcing the halo to be axisymmetric, and aligned with the symmetry axis of the disk. We henceforth explore $c_h$ ($\equiv q_{\rho}$, i.e. the $z$-flattening of the dark matter density distribution) and the circular velocity at the Solar radius $V_{\rm circ} (R_\odot)$. \texttt{MWPotential2014} sets the relative contribution from the bulge, the disk and the halo as $(f_b,f_d, f_h)=(0.05,0.60,0.35)$ that internally normalizes the $V_{\rm circ} (R_\odot=8\kpc)$ to $220\kms$, and hence sets $M_{vir}$ (as well as the total masses of the bulge and disk). In the present work we fix the bulge and disk components to those used in the \texttt{MWPotential2014} model, but we allow the halo component to vary.

The scheme used for the orbit fitting and parameter exploration is then straightforward. Every combination of parameters in the potential model described above can be translated into the corresponding $(V_{circ(R_{\odot})}$ and $q_{\rho})$ values. Therefore, we first grid the $(V_{circ}(R_{\odot}), q_{\rho})$ parameter space ranging between $V_{circ}(R_{\odot})=[210, 270]\kms$ and $q_{\rho}=[0.50, 1.50]$. We grid our parameter space in bins of $3\kms \times 0.05$.

The orbit fitting for a given value of $(V_{\rm circ}(R_\odot), q_{\rho})$ was done as follows. A 6D phase-space starting point is required to integrate an orbit. Without loss of generality, we fixed $\delta = 39\deg$ as a starting point of the orbit (the $\delta = 39\deg$ line passes close to the mid point of the GD-1 stream), and left $\alpha, \overline{\omega}, \mu_{\alpha}, \mu_{\delta}, v_{\rm los}$ as free parameters to be explored by a Markov Chain Monte Carlo (MCMC) algorithm. Every starting 6-D phase-space point was integrated into an orbit that was then compared with the data, defined as sample-1, in order to find the best orbit corresponding to the highest log-likelihood value for the given set of $(V_{\rm circ}(R_\odot), q_{\rho})$. 

To account for contamination that persists in sample-1, we adopted the ``conservative formulation'' of \cite{sivia1996data} which involves a modification of the log-likelihood equation that lowers the contribution from outliers to the likelihood. (We provide a demonstration of the workings of the ``conservative formulation'' in the Appendix). The log-likelihood for each datum $i$ is given by: 
\begin{equation}\label{eq:Loglikehood_Potential1}
\ln \mathcal{L}_i = -\ln ((2\pi)^3\sigma_{\rm sky} \sigma_{\overline{\omega}} \sigma_{G_{BP}-G_{RP}} \sigma_{\mu_{\alpha}} \sigma_{\mu_{\delta}} \sigma_{v_{\rm los}}) +\ln N -\ln D,
\end{equation}
where
\begin{equation}
\begin{aligned}
N &= \prod_{j=1}^6 (1-e^{-R^2_j/2}) \, , \\
D &= \prod_{j=1}^6 R^2_j \, , \\
R_1^2 &= \dfrac{\theta^2_{\rm sky}}{\sigma^2_{\rm sky}} \, , \\
R_2^2 &= \dfrac{(\overline{\omega}_d - (\overline{\omega}_o-0.029))^2}{\sigma^2_{\overline{\omega}}} \, , \\
R_3^2 &= \Big\langle \dfrac{(\,(G_{BP}-G_{RP})_{d} - (G_{BP}-G_{RP})_{o} \,)^2}{\sigma^2_{G_{BP}-G_{RP}}}\, \Big|_{G_k} \Big\rangle \, , \\
R_4^2 &= \dfrac{(\mu_{\alpha, d} - \mu_{\alpha, o})^2}{\sigma^2_{\mu_{\alpha}}} \, , \\
R_5^2 &= \dfrac{(\mu_{\delta, d} - \mu_{\delta, o})^2}{\sigma^2_{\mu_{\delta}}} \, , \\
R_6^2 &= \dfrac{(v_{los,d} - v_{los,o})^2}{\sigma^2_{v_{\rm los}}} \, .\\
 \\
\end{aligned}
\end{equation}
Here, $\theta_{\rm sky}$ is the angular difference between the orbit and the data point, $\overline{\omega}_d, (G_{BP}-G_{RP})_{d}, \mu_{\alpha, d}, \mu_{\delta, d}$  and $v_{los,_d}$ are the observed parallax, color, proper motion and los velocity, with the corresponding orbital model values marked with the subscript ``$o$''. The Gaussian dispersions $\sigma_{\rm sky}, \sigma_{\overline{\omega}}, \sigma_{G_{BP}-G_{RP}}, \sigma_{\mu_{\alpha}}, \sigma_{\mu_{\delta}}, \sigma_{v_{\rm los}}$ are the convolution of the intrinsic dispersion of the model together with the observational uncertainty of each data point. The value of $0.029$ in the parallax term corrects for the zero-point of the parallax measurements present in Gaia DR2 \citep{GaiaDR2_2018_astrometry}. For the $R_3$ term, we calculate the model colour $[G_{BP}-G_{RP}]_{o}$ from the orbital distance and the datum's $G$ magnitude value, by reference to the adopted SSP model mentioned previously in Section \ref{sec:Data}. In order to account for the photometric uncertainties $R_3$ is calculated by averaging over $1000$ randomly-sampled values of $G$ and $(G_{BP}-G_{RP})_{d}$. Finally, the full log-likelihood used in the comparison of the model to the data is then:
\begin{equation}
\ln \mathcal{L}=\sum_i \ln \mathcal{L}_i \, .
\end{equation}

Conversion from Galactocentric coordinates to Heliocentric observables was done by assuming that the Sun is situated at a distance of $R_\odot=8.122\kpc$ from the Galactic centre \citep{GravityCollab2018}, and we set the Sun's peculiar velocity to be $(11.10, 12.24, 7.25)\kms$ \citep{Schornich2010_Sun}.

For every potential $\Phi(x,y,z | V_{\rm circ}(R_\odot), q_{\rho})$, the best fit orbit was found and the corresponding log-likelihood was assigned to the $(V_{\rm circ}(R_\odot), q_{\rho})$ bin. Figure \ref{fig:Orbit_fitting1} presents the resulting contour plot of the parameter exploration and the comparison between the data and the best fit orbit. We find that the likelihood surface is well behaved and peaks at the grid value $(V_{\rm circ}(R_\odot), q_{\rho}) = (\Vcirccross, \qrhocross)$. The best fit values after 1D marginalisation that were obtained are $(V_{\rm circ}(R_\odot), q_{\rho}) = (\Vcirccalc, \qrhocalc)$, thereby placing tight constraints on the circular velocity at the Solar radius and moderate limits on the shape of the dark matter halo assuming this model potential.

\begin{figure}
\begin{center}
\includegraphics[width=\hsize]{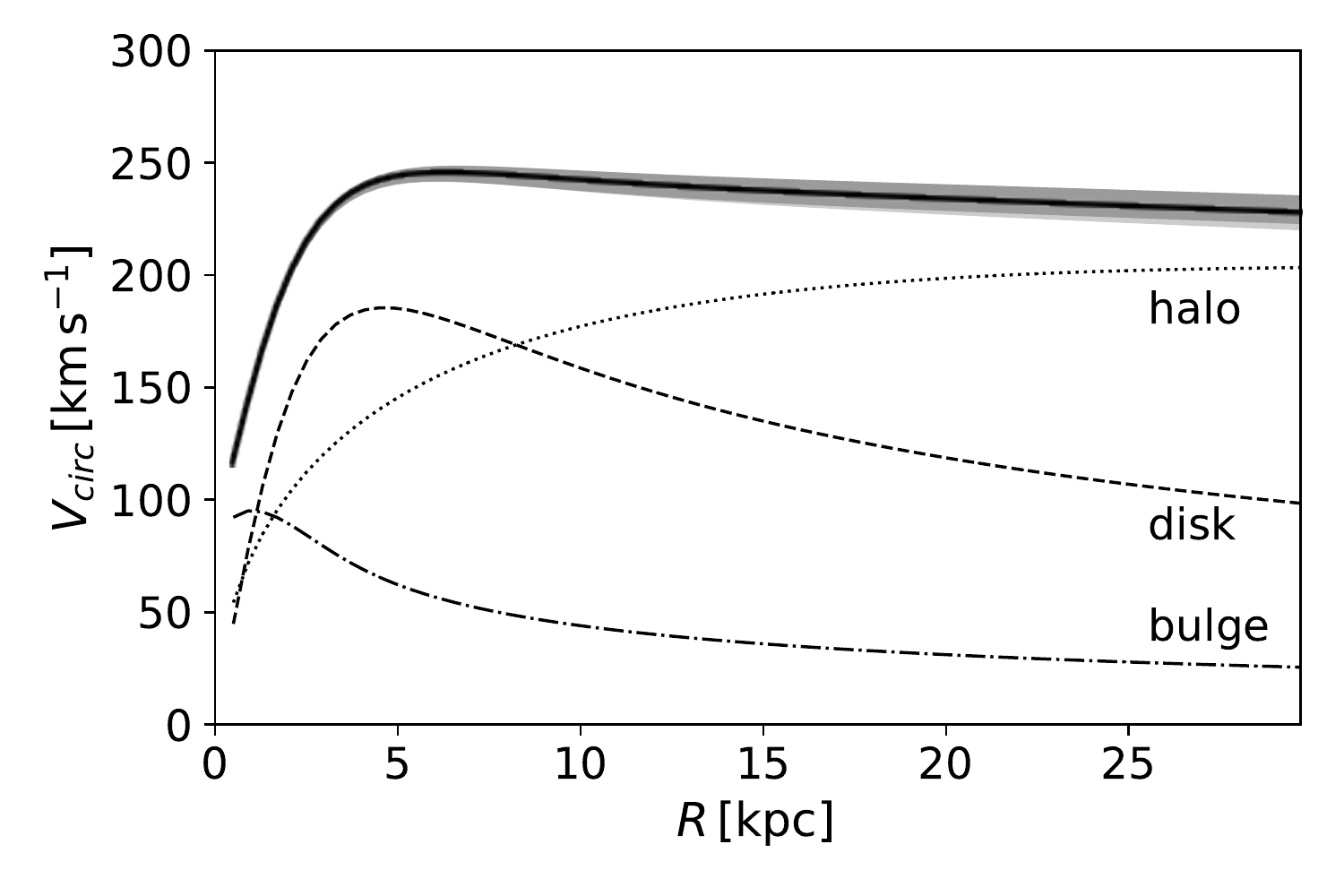}
\vspace{-0.5cm}
\end{center}
\caption{Velocity curve of the Galaxy model discussed in Section \ref{sec:Constrain_MW_potential} incorporating an NFW halo potential. The dashed/dotted curves correspond to the independent velocity curves due to the bulge, the disk and the NFW dark matter halo. The combined circular velocity curve of the Galaxy is plotted in the bold black curve.}
\label{fig:Vel_curve_NFW}
\end{figure}

The resulting Milky Way rotation curve corresponding to $(V_{\rm circ} (R_\odot), q_{\rho}) = (\Vcirccalcwunc, \qrhocalcwunc)$ is shown in Figure \ref{fig:Vel_curve_NFW}. The curve matches expectations for the circular velocity in the outer regions of the Galaxy reasonably well (see Figure 13 of \citealt{Kupper2015} and references therein). The corresponding mass inside of $20\kpc$ (which is well within the orbit of GD-1) is $M_{\rm MW}(R<20\kpc) =\MWmass$.

\begin{figure*}
\begin{center}
\includegraphics[width=0.7\hsize]{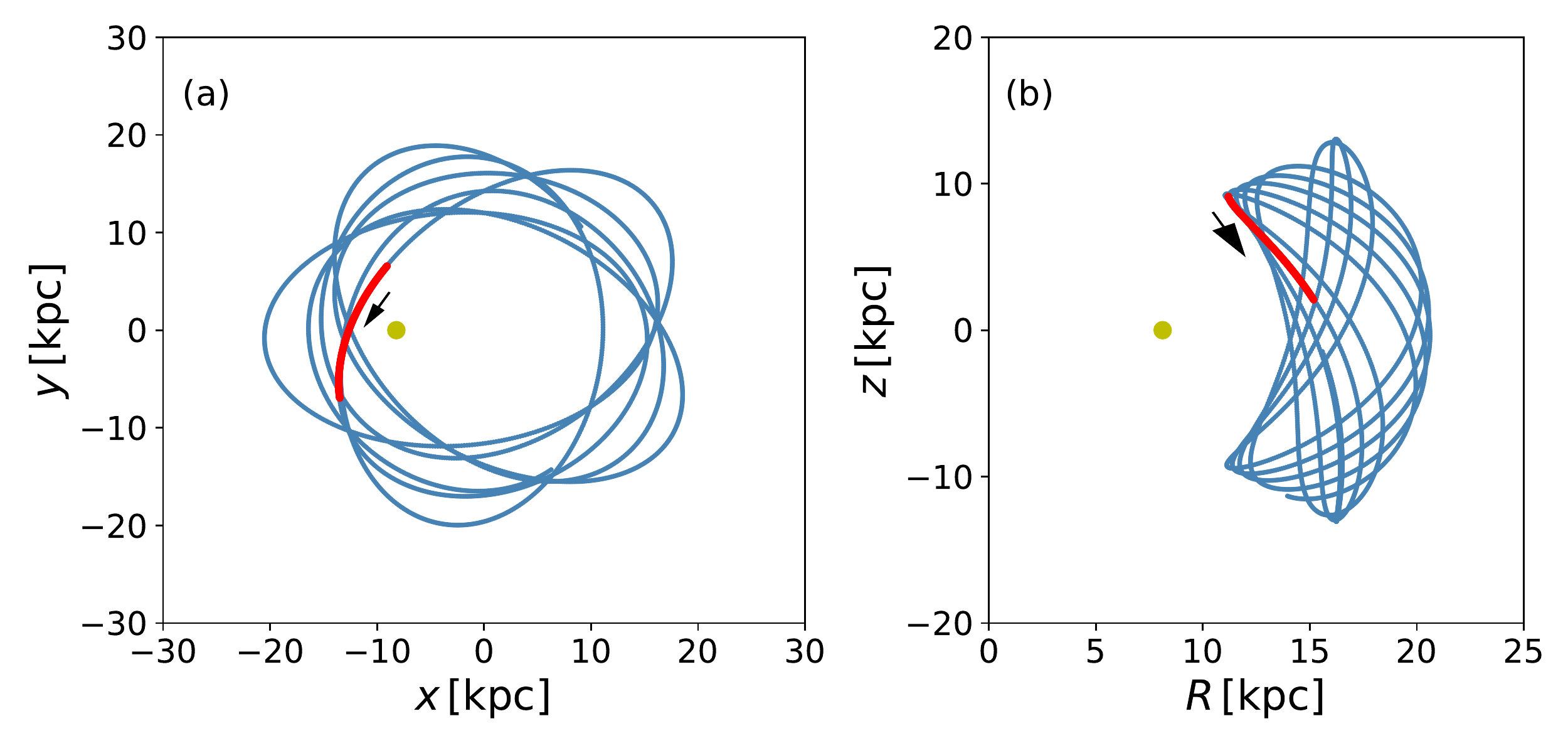}
\end{center}
\vspace{-0.50cm}
\caption{The orbital trajectory of the GD-1 stream, showing the best fit orbit obtained from the orbit-fitting procedure. (a) The orbit (lightblue) is presented in the Galactic $x-y$ plane; for perspective the current location of GD-1 is shown in red. In this Galactocentric Cartesian system the Galactic centre lies at the origin and the Sun (yellow dot) is at $(x,y,z)=(-8.122,0,0.0)\kpc$. The orbit was integrated for $3\Gyr$. (b) Same as (a) but in the Galactic $R-z$ plane. The arrows represent the direction of motion of GD-1. For this orbit we found $(\rm{r_{apo},r_{peri},eccentricity)} = (\rapo \kpc, \rperi \kpc, \ecc)$.}
\label{fig:orbit_behaviour_plot}
\end{figure*}

The orbital trajectory of the best-fit orbit is shown in Figure \ref{fig:orbit_behaviour_plot}, integrated over a period of $3\Gyr$ in the best-fit potential model. The orbit of GD-1 appears to be loop-like and is strongly retrograde, possessing an apocenter at $r_{\rm \rm apo}=\rapo \kpc$, a pericenter at $r_{peri}=\rperi \kpc$, a maximum height from the Galactic plane of  $z_{\rm max}=\zmax \kpc$ and an eccentricity of $e=\ecc$ (these values are also tabulated in Table \ref{tab:GD1_properties}).

\begin{table}
\caption{Properties of the GD-1 stellar stream.}
\label{tab:GD1_properties}
\begin{center}
\begin{tabular}{cc}
\hline
$\rm{Parameter}$ & $\rm{Range/Value}$\\
\hline
R.A. & $[130\deg, 220\deg]$\\
Dec & $[20\deg, 60\deg]$\\
$D_{\odot}(\kpc)$ & $[7, 12]$\\
$\sigma_w$(pc) & $50$\\
$\mu^{*}_{\alpha}(\masyr)$ & $[-9, -2]$\\
$\mu_{\delta}(\masyr)$& $[-14, -1]$\\
$v_{los}(\kms)$ & $[-300, 250]$\\
$\sigma_{v}(2D, \kms)$ & $\veldispunitless$ (95\% conf.)\\
$\rm{[Fe/H]}$(dex) & $\FeHcalc$\\
$\rm{z_{max}}(\kpc)$ & $\zmax$\\
$\rm{r_{peri}}(\kpc)$ & $\rperi$\\
$\rm{r_{apo}}(\kpc)$ & $\rapo$\\
e & $\ecc$\\
$L_z(\kpc\kms)$ & $\Lzang$\\
\hline
  
\end{tabular}
\end{center}
\end{table}

\begin{figure*}
\begin{center}
\vbox{
\includegraphics[width=0.8\hsize]{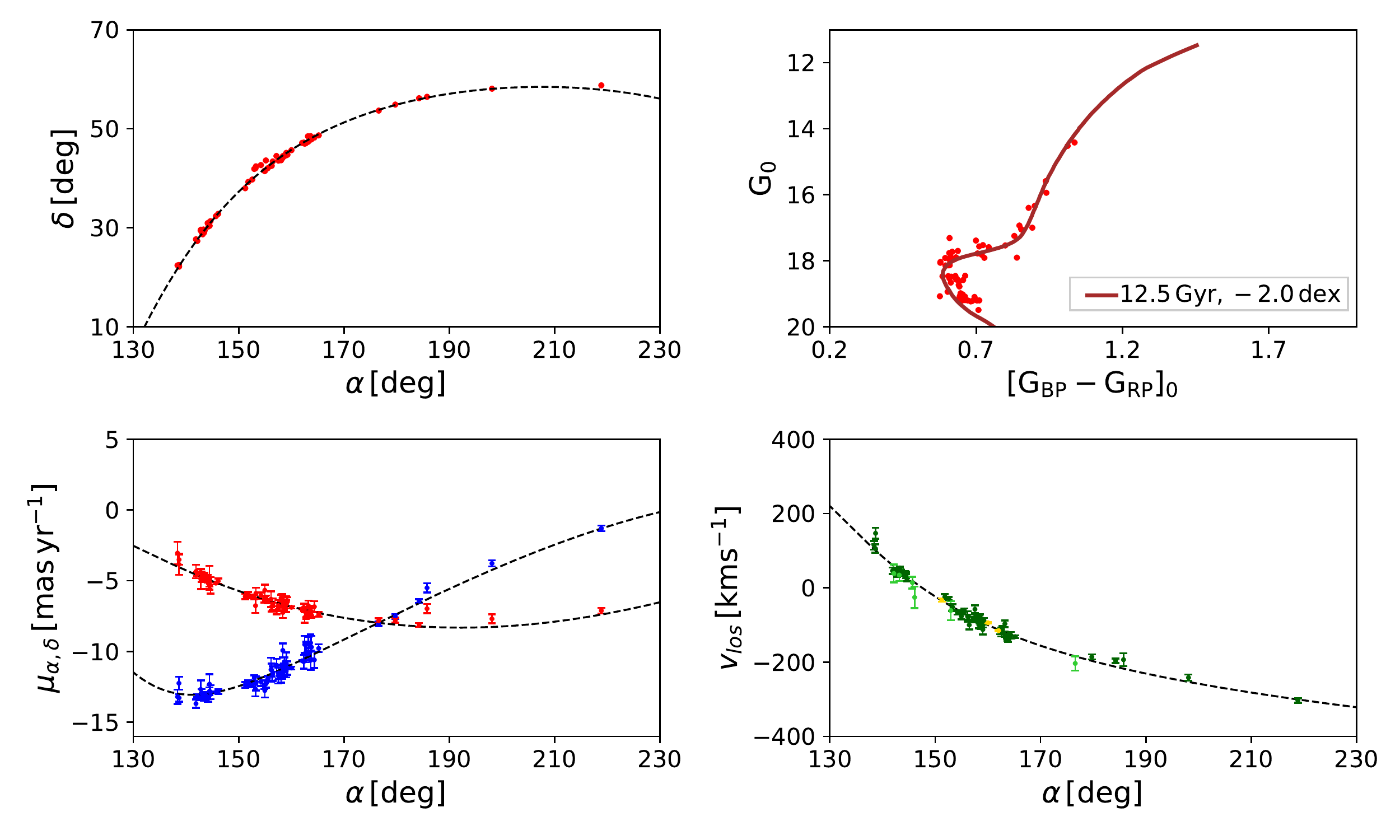}
}
\hbox{
\hspace{1.5cm}
\includegraphics[width=0.28\hsize]{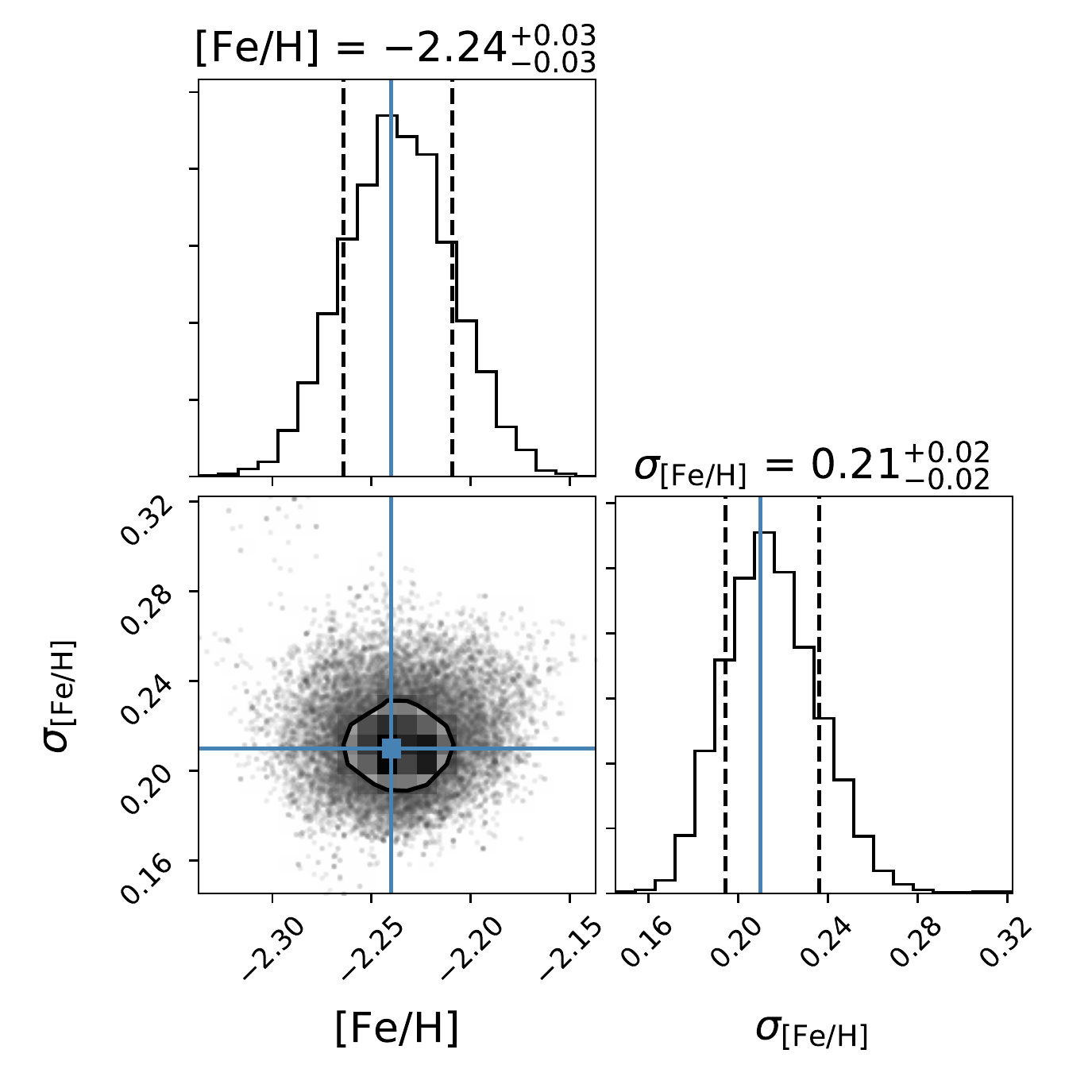}
\hspace{0.50cm}
\includegraphics[width=0.5\hsize]{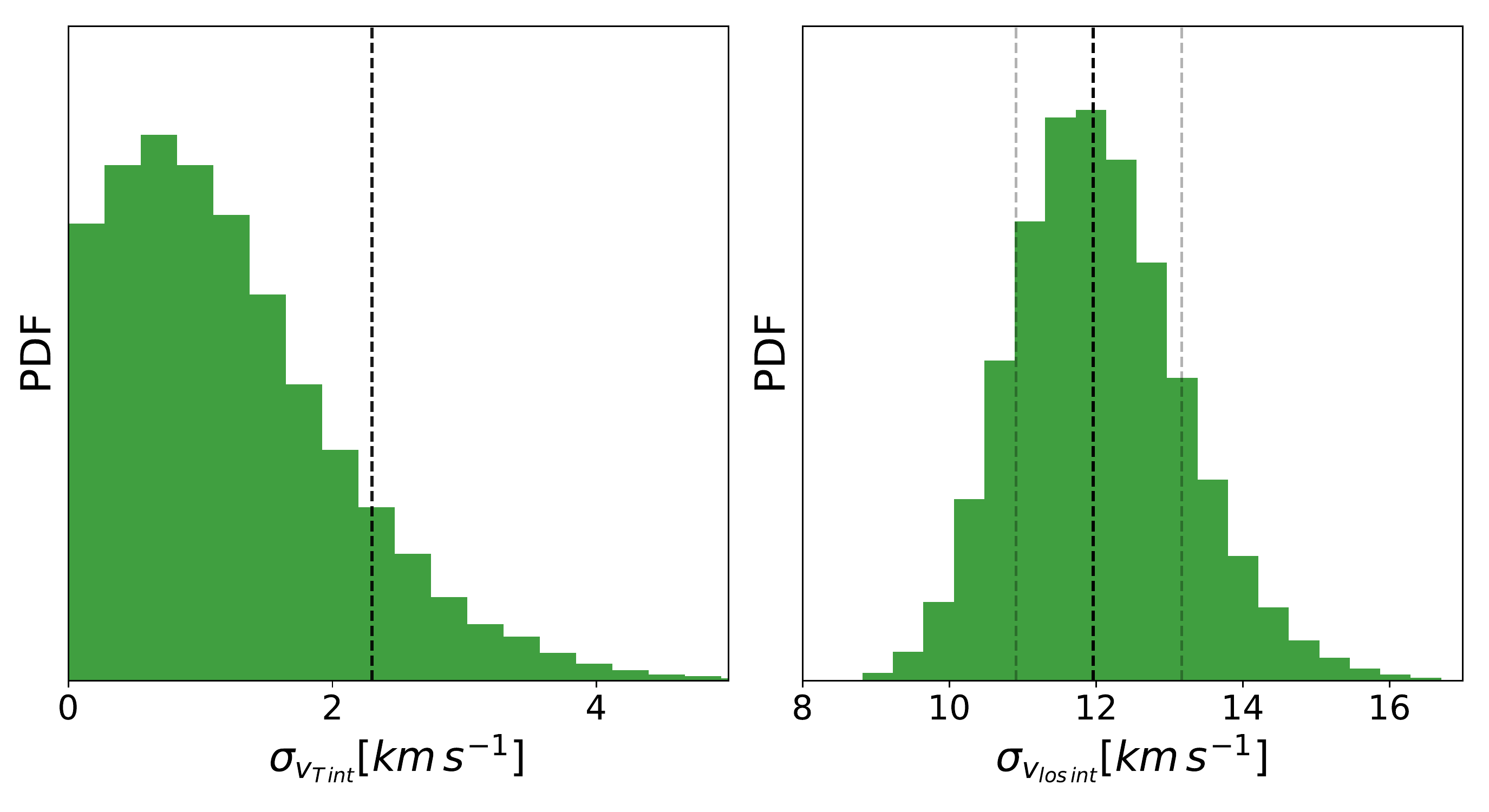}
}
\vspace{-0.5cm}
\end{center}
\caption{GD-1's physical and chemical properties : 
{\it Top panels:} High confidence GD-1 members obtained by sigma clipping data points from sample-1 based on the best fit orbital model obtained by our analysis presented in Section \ref{sec:Constrain_MW_potential}. We refer to this sample as sample-2.
{\it Bottom left panels:} PDF for the mean metallicity $(\rm{[Fe/H]})$ of the GD-1 stream and the corresponding metallicity dispersion $(\sigma_{\rm{[Fe/H]}})$. 
{\it Bottom right panels:} Velocity dispersion of the GD-1 stream along the tangential (left) and line of sight (right) directions. In the left panel the dashed line indicates the 95\% confidence upper limit on $\sigma_{v_T \, \rm int}$, whereas on the right they indicate $1\sigma$ limits.}
\label{fig:vel_dispersion_feh}
\end{figure*}

\section{Velocity dispersion and Metallicity of GD-1}
\label{sec:GD1_vel_Feh_Feh}

In this Section we estimate the velocity dispersion and the metallicty of the GD-1 structure. To do this, we first make use of the analysis performed in Section \ref{sec:Constrain_MW_potential} to improve the sample selection. The best fit orbit, shown in Figure \ref{fig:Orbit_fitting1}, passes through those stars that have higher likelihood of being GD-1 members. We use this orbit model to undertake a sigma clipping procedure in order to select only those stars in sample-1 that lie within $5\sigma$ of the model in any of the observed phase-space parameters. This selection results in $67$ stars (out of the $80$ stars in sample-1). We refer to this new sample as sample-2 (shown in Figure \ref{fig:vel_dispersion_feh}). With this sample of high confidence GD-1 stars, we can employ the usual log-likelihood expression to measure the physical attributes of GD-1 that are of interest for this study. 

\subsection{Velocity dispersion}
The very fine pencil-line track of GD-1 suggests that the stream must be dynamically cold and hence is possibly a remnant of some (possibly completely disrupted) globular cluster. We test this hypothesis here by measuring the velocity dispersion of the stream.

For an isotropic system, the internal velocity dispersion $\sigma_{\rm int}$ can be expressed as sum of its components as:
\begin{equation}\label{eq:vel_disp}
\sigma^2_{\rm int}=\sigma^2_{v_T \, \rm int} +\sigma^2_{v_{\rm los} \, \rm int}\,,
\end{equation}
where $\sigma_{v_{T} \, \rm int}$ and $\sigma_{v_{\rm los} \, \rm int}$ are, respectively, the tangential and the radial components of the velocity dispersion. In order to fully exploit Gaia's precise proper motion measurements, we decided to estimate $\sigma_{v_T \, \rm int}$ and $\sigma_{v_{\rm los} \, \rm int}$ independently. This was also done to ensure that the relatively large uncertainties in the line of sight velocity measurements remain separated from Gaia's precisely-measured proper motions. We use sample-2 as our data and the best-fit orbit (obtained in Section \ref{sec:Constrain_MW_potential}) as our model. The log-likelihood functions are taken to be:
\begin{equation}\label{eq:Vel_disp_no_contam}
\begin{split}
\ln \mathcal{L}_1 &= \sum_{\rm data}  -\ln(\sigma_{v_T \, \rm obs})-\frac{1}{2} \Big(\dfrac{v^m_T-v^d_T}{\sigma_{v_T \, \rm obs}}\Big)^2\\
\ln \mathcal{L}_2 &= \sum_{\rm data}  -\ln(\sigma_{v_{\rm los} \, \rm obs})- \frac{1}{2} \Big(\dfrac{v^m_{\rm los}-v^d_{\rm los}}{\sigma_{v_{\rm los} \, \rm obs}}\Big)^2\,,
\end{split}
\end{equation}
where $v^d_T$ is the observed tangential velocity of the data calculated by multiplying the orbit model distance with the proper motion measurement, and $v^d_{\rm los}$ is the observed los velocity. The corresponding orbital model values are marked with superscript `m'. The Gaussian dispersions $\sigma_{v_T \, \rm obs}$ and $\sigma_{v_{\rm los} \, \rm obs}$ are the convolution of the intrinsic dispersion of the model together with the observational uncertainty of each data point ($\sigma^2_{obs}=\sigma^2_{\rm int} + \delta^2_i$, with $\delta_i$ being the measured uncertainty of the data). 

A Markov chain Monte Carlo algorithm was used to survey the parameter space of $\sigma_{v_T \, \rm int}$ and $\sigma_{v_{\rm los} \, \rm int}$. The resulting distribution is shown in Figure \ref{fig:vel_dispersion_feh}. In the direction tangential to the line of sight, we find $\sigma_{v_T \, \rm int} <\veldisp$ (at the 95\% confidence level), whereas in the line of sight direction, we obtain $\sigma_{v_{\rm los} \, \rm int} = 11.95^{+1.20}_{-1.05}\kms$. 

The value of $\sigma_{v_T \, \rm int}$ clearly shows that the GD-1 stream system is dynamically extremely cold, which is consistent with it being a remnant of some very low mass system, such as a globular cluster. The relatively higher value of $\sigma_{v_{\rm los} \, \rm int}$ suggests that the observational uncertainties of the stars in the radial velocity surveys are underestimated, but note that the average los velocity uncertainty is $\sim8\kms$, which greatly exceeds the internal velocity dispersion tangential to the line of sight. Admittedly, the high value of velocity dispersion could also be due to the plausible radial velocity offset in the LAMOST dataset with respect to the SEGUE datset (a similar offset that has been recently reported between the APOGEE DR14 and LAMOST DR3 surveys, \citealt{2018A&A...620A..76A}) The value for $\sigma_{v_T \, \rm int}$ obtained here for GD-1 could play a crucial role for future N-body dynamical modeling of the stream and for assessing the impact of any perturbers (such as the dark matter sub-halos) on this structure \citep{Bonaca_spur_2018}.

\subsection{Metallicity}
Equipped with the metallicty measurements from SEGUE and LAMOST, we now calculated the metallicity of GD-1. We ran an MCMC algorithm on the data of sample-2, and found $\rm{[Fe/H]}=\FeHcalc$ (see Figure \ref{fig:vel_dispersion_feh}). This low value of metallicity makes GD-1 an extremely metal poor halo substructure (see \cite{GrillmairCarlin2016} for comparison with other streams). 

We also note that this value of the system's metallicity is similar to the ${\rm [Fe/H]}$ of the adopted SSP template model that we have used hitherto for our GD-1 analysis. Although the differences are minor, the adopted SSP model follows the {\it photometric} behaviour of the GD-1 stars slightly better.

\section{Discussion and conclusions}\label{sec:Conclusions}

In this contribution we probe the underlying gravitational potential of the Milky Way by fitting the long $(>70\deg)$ and narrow GD-1 stellar stream with a realistic family of Galaxy models, changing the circular velocity at the Solar radius $(V_{\rm circ}(R_\odot))$ and the shape of the dark matter halo $(q_{\rho})$. 

For this, GD-1 members were first recovered from our \texttt{STREAMFINDER} density map (Figure \ref{fig:Streamfinder_density_plot}) that was obtained from processing the Gaia DR2 catalogue\citep{Malhan2018_SF, Malhan_Ghostly_2018, Ibata_Norse_streams2019}. This GD-1 dataset was then cross-matched with the SEGUE and LAMOST data, to acquire stellar $v_{\rm los}$ and [Fe/H] information that is missing from Gaia DR2 (Figure \ref{fig:Streamfinder_GD1_stars_raw}, these stars are also listed in Table \ref{tab:GD1_inventory}). 

The gravitational potential model we studied possesses a universal model for the dark matter halo together with reasonable models for the stellar bulge and disk components. The fitted properties of this Galaxy model imply a circular velocity at the Solar radius and a halo flattening of $(V_{\rm circ}(R_\odot), q_{\rho}) = (\Vcirccalc, \qrhocalc)$  (see Figure \ref{fig:Orbit_fitting1}).

This estimate of $V_{\rm circ} (R_\odot)$ is in good agreement with the estimate of  \cite{Kupper2015} where they obtained $V_{\rm circ} (R_\odot=8.30\kpc)=243\pm16\kms$ from their dynamical study of the Palomar~5 stream. Our estimate of  $V_{\rm circ}(R_\odot)$ is in excellent agreement with those obtained by other authors based on different  approaches. For  example,  \cite{McMillan2011MWMassModel} used photometric and kinematic data to fit  mass models of the Milky Way and found $V_{\rm circ} (R_\odot=8.29\kpc)=239\pm5\kms$;  \cite{Reid2014_Sun} estimated $V_{\rm circ} (R_\odot=8.34\kpc)=240\pm8\kms$ from studies based on the dynamics of the high-mass star forming regions in the spiral arms of the Milky Way. Slightly lower values of this parameter were found by \cite{Koposov2010} $(V_{\rm circ}(R_\odot)=221^{+16}_{-20}\kms)$ and \cite{Sgr_reflex} $(V_{\rm circ}(R_\odot)=229\pm6\kms)$ from an analysis of the orbit of GD-1 and from the geometry of the Sagittarius stream, respectively. It is in stronger tension with, the value recently obtained by \cite{Eilers2018VelCurve} $(V_{\rm circ}(R_\odot)=229.0\pm0.2\kms)$ from an analysis of the velocity curve of the Milky Way combining data from APOGEE, WISE, 2MASS and Gaia.

As for $q_{\rho}$, which was only modestly constrained in our study, our result is consistent with the joint analysis of the kinematics of the GD-1 and Palomar~5 streams by \citet{Bovy2016GD1Pal5} ($q_{\rho}=0.86\pm 0.04$). Our value is also fairly compatible with the recent measurement by \cite{Posti2018MWMass} who find prolate halo solutions for the Milky Way halo with $q_{\rho}=1.3\pm0.25$, based on an analysis of the globular clusters. 

With our model we estimate the mass of the Milky Way in the inner $20\kpc$ (which is still well within the orbit of GD-1, Figure~\ref{fig:orbit_behaviour_plot}) to be $M_{\rm MW} (<20\kpc) =\MWmass$. This value is similar to the value derived from an analysis of globular cluster motions in Gaia DR2 by \citet{Watkins2018_GCanalysis} of $M_{\rm MW}(<21.1\kpc)=2.2^{+0.4}_{-0.3} \times 10^{11}\msun$ and is comparable with the analysis of \citet{Kupper2015}, who found $M_{\rm MW}(<19\kpc)=2.1\pm 0.4 \times 10^{11}\msun$ from the phase-space structure of the Palomar~5 stream. Our value is also compatible with the findings of \cite{Posti2018MWMass}, who obtained $M_{\rm MW}(<20\kpc)=1.91^{+0.17}_{-0.15} \times 10^{11}\msun$. 

The agreement between these studies with different approaches and different dynamical tracers suggests that the mass in the inner regions of the halo is beginning to be understood, although the extra-planar distribution (i.e. what is often modelled as an ellipsoidal ``flattening'') is still quite uncertain. Nevertheless, these results are dependent on the models and associated parameters that have been assumed in the various studies, and in particular the corresponding uncertainties have to be interpreted with care.

We also used the well-measured Gaia proper motions to estimate the internal velocity dispersion of the GD-1 stream stars. Although we could not put tight constraints on the line of sight dispersion component due to the large uncertainties in the SEGUE and LAMOST line of sight velocity measurements, we could place strong limits on the (2D) velocity dispersion tangential to the line of sight, which we find to be $\sigma_{v_T \, \rm int} < \veldisp$ at the $95\%$ confidence level (Figure \ref{fig:vel_dispersion_feh}). This limit corresponds to a 1D velocity dispersion value of $\sim 1.5\kms$. In addition to indicating that GD-1 is an extremely dynamically cold system and indeed the remnant of a globular cluster, such a low velocity dispersion also suggests that so far GD-1 has not suffered substantial external heating, due to interactions with the disk, bar, or any halo substructures in the Milky Way (such as the dark matter sub-halos). 

The new GD-1 inventory of high-likelihood GD-1 members also allowed us to analyse the chemistry of the GD-1 structure. Using metallicity measurements from SEGUE and LAMOST, we determined GD-1's metallicity to be $\rm{[Fe/H]=\FeHcalc}$ (Figure \ref{fig:vel_dispersion_feh}).

Thanks to Gaia's remarkably precise proper motion measurements, we were able to obtain reasonable constraints on the Milky Way's $(V_{\rm circ}(R_\odot), q_{\rho})$ parameters by analysing only a single stream structure. However, the solutions are model-dependent, and so it will be useful to readdress this problem with improved Milky Way models once the mass distribution in the disk and bulge are better constrained from future Gaia studies. A further caveat is that our analysis is based on the assumption that GD-1 perfectly delineates an orbit through the Galaxy; this is only an approximation, and the influence of the assumption should be re-assessed with N-body simulations once the position of the progenitor remnant is securely known. It is likely that armed with Gaia's unprecedentedly accurate proper motions, performing similar analyses with an ensemble of streams will ultimately unleash the full power of tidal streams, possibly providing much improved constraints on the underlying potential and dark matter density of the Milky Way halo, that can then be extrapolated out to larger Galactic radii with more confidence.

\section*{ACKNOWLEDGEMENTS}
The authors would like to acknowledge the comments from the anonymous reviewer. KM acknowledges  support  by the $\rm{Vetenskapsr\mathring{a}de}$t (Swedish Research Council) through contract No.  638-2013-8993 and the Oskar Klein Centre for Cosmoparticle Physics.

This work has made extensive use of \texttt{galpy} module \citep{BovyGalpy2015} for setting up the different galactic potential models and for the purpose of orbit integration.

This work has made use of data from the European Space Agency (ESA) mission {\it Gaia} (\url{https://www.cosmos.esa.int/gaia}), processed by the {\it Gaia} Data Processing and Analysis Consortium (DPAC, \url{https://www.cosmos.esa.int/web/gaia/dpac/consortium}). Funding for the DPAC has been provided by national institutions, in particular the institutions participating in the {\it Gaia} Multilateral Agreement. 

Guoshoujing Telescope (the Large Sky Area Multi-Object Fiber Spectroscopic Telescope LAMOST) is a National Major Scientific Project built by the Chinese Academy of Sciences. Funding for the project has been provided by the National Development and Reform Commission. LAMOST is operated and managed by the National Astronomical Observatories, Chinese Academy of Sciences. 

Funding for SDSS-III has been provided by the Alfred P. Sloan Foundation, the Participating Institutions, the National Science Foundation, and the U.S. Department of Energy Office of Science. The SDSS-III web site is \url{http://www.sdss3.org/}.

SDSS-III is managed by the Astrophysical Research Consortium for the Participating Institutions of the SDSS-III Collaboration including the University of Arizona, the Brazilian Participation Group, Brookhaven National Laboratory, Carnegie Mellon University, University of Florida, the French Participation Group, the German Participation Group, Harvard University, the Instituto de Astrofisica de Canarias, the Michigan State/Notre Dame/JINA Participation Group, Johns Hopkins University, Lawrence Berkeley National Laboratory, Max Planck Institute for Astrophysics, Max Planck Institute for Extraterrestrial Physics, New Mexico State University, New York University, Ohio State University, Pennsylvania State University, University of Portsmouth, Princeton University, the Spanish Participation Group, University of Tokyo, University of Utah, Vanderbilt University, University of Virginia, University of Washington, and Yale University.

\clearpage

\begin{table*}
\caption{GD-1 star members with measured line-of-sight velocities from SEGUE and LAMOST (only the first 10 entries are listed, the full table is available in electronic format). The sky coordinates and photometry information are provided by Gaia DR2, where the latter is  corrected for extinction. The los velocities and metallcities are measurements from SEGUE and LAMOST which are referred to, respectively, as S or L in the Source column  (stars that are both in SEGUE and LAMOST are marked $S \cap L$. The corresponding measurements are weighted averages).}
\label{tab:GD1_inventory}
\begin{center}
\begin{tabular}{cccccccccc}
\hline
RA J2000 & Dec J2000 &  ${\rm G_0}$ & $[G_{\rm BP} -G_{\rm RP}]_0$ & $\rm{v_{los}}$ & $\delta_{\rm v_{los}}$ & [Fe/H] & $\delta_{\rm[Fe/H]}$ & Source & Probable\\

[deg] & [deg] & [mag] & [mag] & $[\kms]$ & $[\kms]$  & [dex] & [dex] & & member?\\
\hline

138.39986 & 22.41298 & 19.23 & 0.68 & 114.52 & 11.59 & -2.47 & 0.05 & S & Y \\
138.45075 & 22.43453 & 17.75 & 0.72 & 138.31 & 4.05 & -2.16 & 0.1 & S & N \\
138.66816 & 22.4817 & 19.2 & 0.65 & 105.69 & 11.13 & -1.77 & 0.03 & S & Y \\
138.68803 & 22.15384 & 18.14 & 0.6 & 146.86 & 14.52 & -9999.0 & -9999.0 & S & Y \\
141.11201 & 27.5518 & 16.72 & 0.71 & 154.9 & 12.01 & -1.11 & 0.26 & L & N \\
141.65383 & 29.35077 & 12.9 & 1.1 & 28.2 & 6.59 & -1.49 & 0.09 & L & N \\
141.89885 & 27.68611 & 18.54 & 0.63 & 45.87 & 8.67 & -2.28 & 0.08 & S & Y \\
142.15018 & 27.3378 & 16.4 & 0.88 & 38.63 & 23.99 & -2.05 & 0.38 & L & Y \\
142.78781 & 29.46105 & 17.06 & 0.86 & 47.81 & 4.16 & -2.05 & 0.04 & S & Y \\
142.82166 & 29.61698 & 19.19 & 0.66 & 37.02 & 7.27 & -2.36 & 0.06 & S & Y \\
151.24928 & 37.99899 & 17.04 & 0.85 & -33.15 & 4.71 & -2.32 & 0.08 & $S \cap L$ & Y \\

\hline
\end{tabular}
\end{center}
\end{table*}

\section*{Appendix 1}\label{sec:Appendix1}

In the study presented here, we employed the ``conservative formulation'' of \cite{sivia1996data} for calculating the log-likelihood functions (presented in Section \ref{sec:Constrain_MW_potential}). This method lowers the influence of outliers in a sample and thereby provides a means to account for contamination without having to develop an explicit model of the contamination (which can often be very difficult to devise). The method is a generalization of standard least squares, where one assumes that the uncertainty estimates $\sigma_0$ that have been measured are in reality only a lower bound on the actual uncertainty $\sigma$. The probability density function for this uncertainty is taken to be $P(\sigma) = \sigma_0/\sigma^2$ for $\sigma >\sigma_0$ (otherwise, $P(\sigma) = 0$).

In this section, we demonstrate the power of this new method over the standard likelihood through a simple example. We draw this comparison by analyzing two types of fiducial datasets - one that is completely devoid of contamination and the other infused with contaminants. 

\begin{figure}
\begin{center}
\vbox{
\includegraphics[width=0.60\hsize]{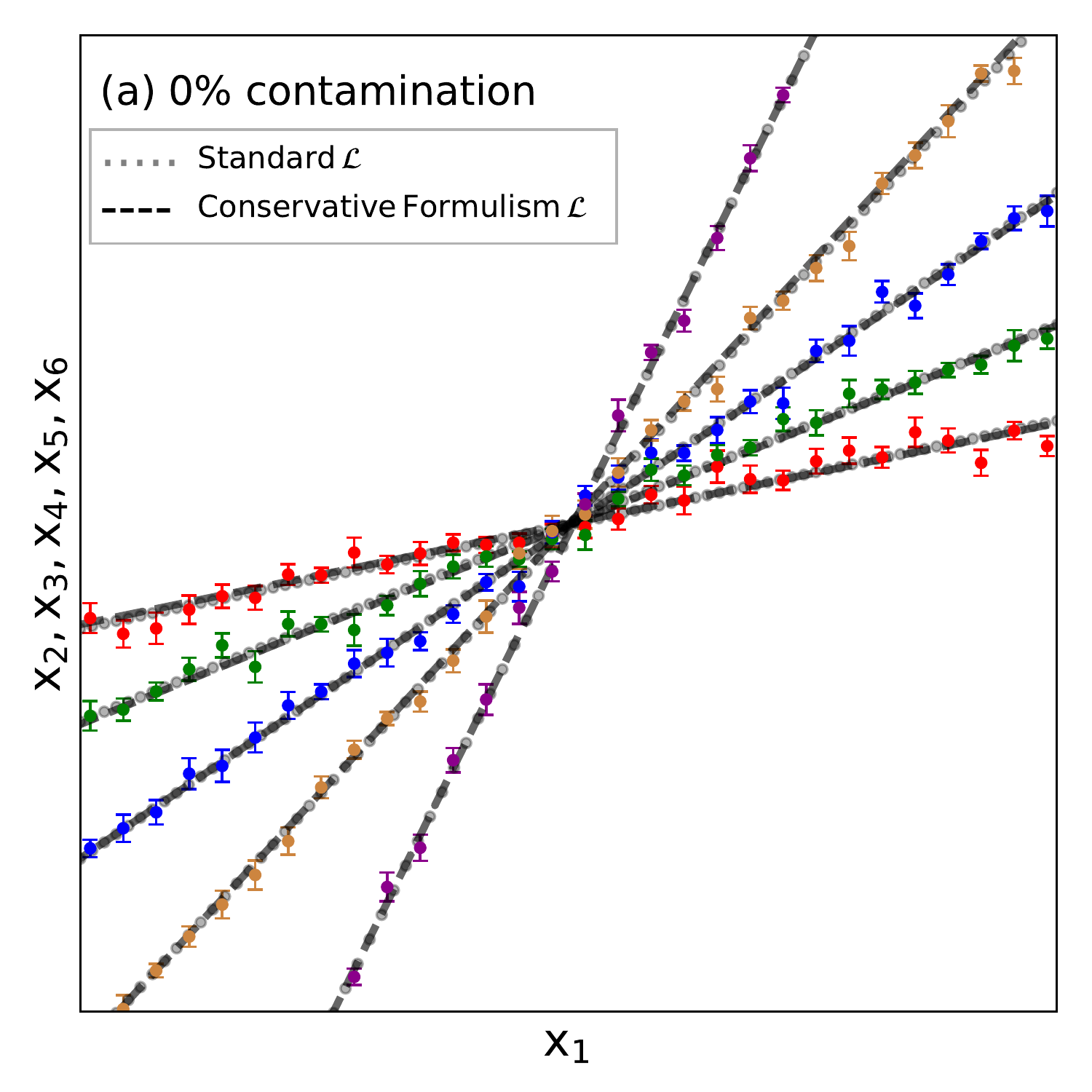}
\includegraphics[width=0.60\hsize]{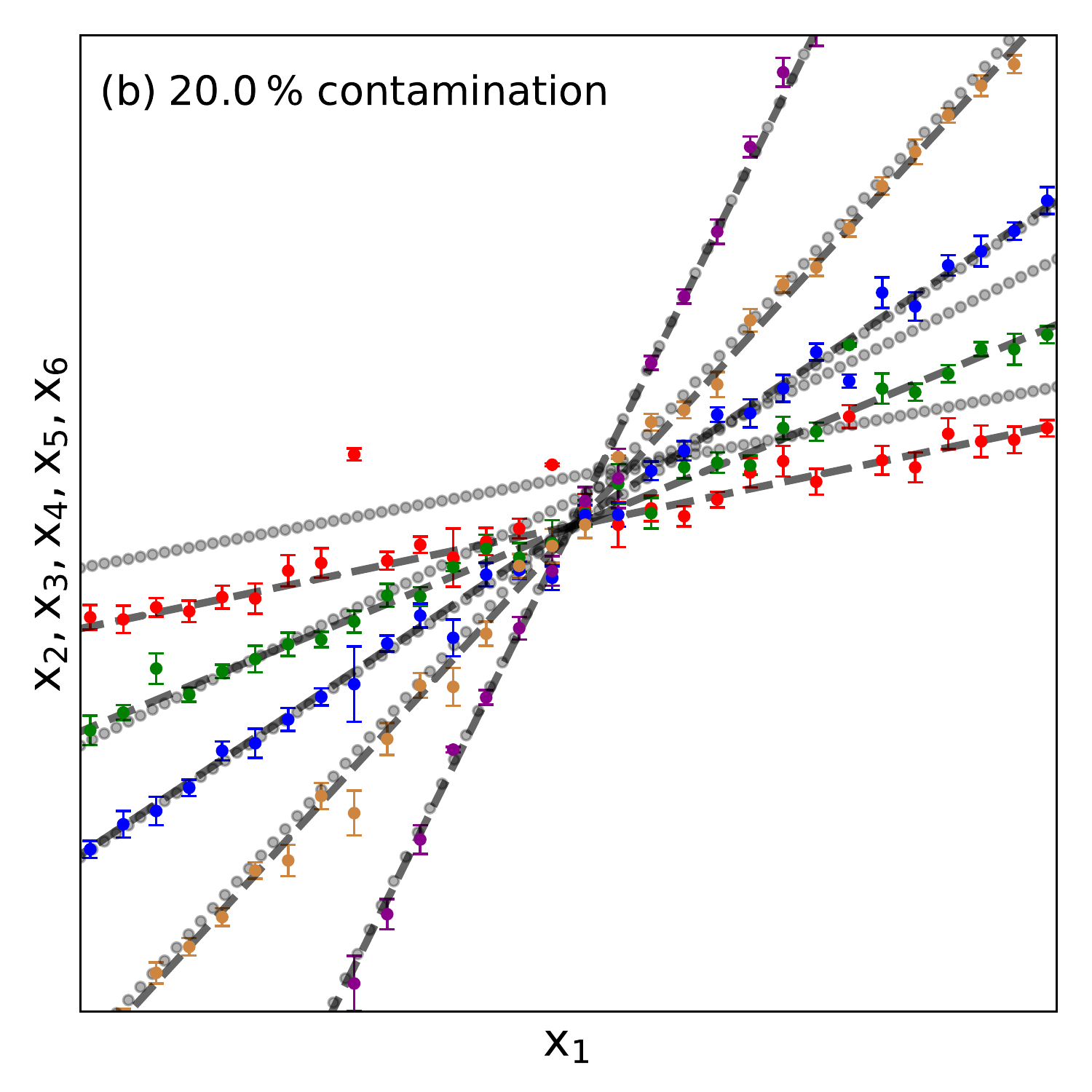}
}
\end{center}
\caption{Comparison between the workings of the conservative formulation log-likelihood function and the standard likelihood function. (a) Fiducial data for a 6D straight line. The X-axis of the plot corresponds to one of the dimensions $(x_1)$, and the Y-axis represents the remaining dimensions $(x_2, x_3, x_4, x_5, x_6)$ (shown in different colours). The dotted and the dashed lines correspond to the solutions obtained by employing log-likelihood functions described by equations \ref{eq:Loglikehood_standard} and \ref{eq:Loglikehood_Sivia} respectively. (b) Same as (a), but with $20$\% contamination.}
\label{fig:SiviaDemo}
\end{figure}

For this, we constructed a perfect 6D straight line and then made it noisy by introducing random uncertainties. This dataset is shown in Figure \ref{fig:SiviaDemo}a. We first fit this data with the standard likelihood definition, which for each datum $i$ is expressed as:

\begin{equation}\label{eq:Loglikehood_standard}
\ln \mathcal{L}_i = -\ln (\prod_{j=1}^5 \sigma_{i,j}) -\dfrac{1}{2} (\sum_{j=1}^5 X^2_{i,j})
\end{equation}
where
\begin{equation}
\begin{aligned}
X^2_{i,j}=\Big(\dfrac{x_{i,j}^d -x_{i,j}^m}{\sigma_{i,j}}\Big)^2 \, .
\end{aligned}
\end{equation}
Here, the product and summation are done over all the dimensions of the line. The data values and the model values are marked with $d$ and $m$ superscripts, respectively. $\sigma_{i,j}$ refers to the Gaussian dispersion which is a convolution of the intrinsic dispersion of the model together with the data uncertainty for datum $i$ in the $\rm{j^{th}}$ dimension ($\sigma^2_i = \sigma^2_{\rm int} +\delta^2_i$, with $\delta_i$ being the uncertainty of the data). Finally, the full log-likelihood used in the comparison of the model to the data is then:
\begin{equation}
\ln \mathcal{L}=\sum_i \ln \mathcal{L}_i \, .
\end{equation}

Next, we fit the same data but using the ``conservative formulation'', which for each datum $i$ is expressed as:
\begin{equation}\label{eq:Loglikehood_Sivia}
\ln \mathcal{L}_i = -\ln ((2\pi)^{5/2} \prod_{j=1}^5 \sigma_{i,j}) +\ln N_i -\ln D_i,
\end{equation}
where
\begin{equation}
\begin{aligned}
N_i &= \prod_{j=1}^5 (1-e^{-X^2_{i,j}/2}) \, , \\
D_i &= \prod_{j=1}^5 X^2_{i,j} \, . \\
\end{aligned}
\end{equation}
The best fit straight line solutions for both the cases are shown in Figure \ref{fig:SiviaDemo}a. The two best fit lines, obtained via two different likelihood formulations, overlay each other (and are also close to the true solution). The \cite{sivia1996data} method performs slightly worse, recovering parameters values with $\sim 2$\% error in this case, compared to $\sim 1$\% for the standard likelihood method.

We then introduce a $20\%$ level of contamination into this fiducial dataset, forcing some of the datapoints to become outliers. This contaminated data is shown in Figure \ref{fig:SiviaDemo}b. Some of the contaminants can be easily seen in this plot. The resulting best fit straight line solutions in this case can be seen in Figure \ref{fig:SiviaDemo}b. The solution corresponding to the ``conservative formulation'' approach fits is much less affected by the outliers, as it recovers the input parameters to within $\sim 4$\%, whereas the standard likelihood method gives $\sim 15$\% errors.

This simple example demonstrates that the ``conservative formulation'' of \cite{sivia1996data} provides an effective means to de-contaminate a sample with multi-dimensional properties in an objective way.

\bibliographystyle{mnras}
\bibliography{main} 

\bsp	
\label{lastpage}
\end{document}